\pdfoutput=1
\documentclass[12pt]{article}
\usepackage{graphicx}
\usepackage{amsmath}
\usepackage{amssymb}
\usepackage{gensymb}
\usepackage{dcolumn}
\def\be{\begin{equation}}
\def\ee{\end{equation}}
\def\ba{\begin{array}}
\def\ea{\end{array}}
\def\bpm{\begin{pmatrix}}
\def\epm{\end{pmatrix}}
\def\beqn{\begin{eqnarray}}
\def\eeqn{\end{eqnarray}}
\def\nonum{\nonumber}
\def\bt{\begin{tabular}}
\def\et{\end{tabular}}
\def\bc{\begin{center}}
\def\ec{\end{center}}

\begin{document}
\title{Minimizing the phase structure of quark mass matrices} 
\author { Nikhila Awasthi~$^1$, Manoj Kumar~$^2$, Monika Randhawa~$^2$ and Manmohan Gupta~$^1$
\\
{\small {\it 
$^1$ Department of Physics, Centre of Advanced Study, Panjab University,}}\\
{\small {\it Chandigarh, 160014, India.}}\\
{\small {\it $^2$ University Institute of Engineering and Technology, Panjab
University,}} \\
{\small {\it  Chandigarh, 160014, India.}}}
 \maketitle
 \begin{abstract}
Fritzsch-Xing matrices  are a particular class of texture 4 zero hermitian quark mass matrices, known to be successful in accommodating  the  quark mixing data. In the present work,  it is shown that these  texture 4-zero matrices with only one phase parameter, unlike the usually  considered two phase parameters,   are  not only consistent with the latest experimental quark mixing  data, but  also predict the CP violation parameters, $J$ and corresponding phase $\delta$,  in agreement with the recent global analyses. We also show that the mass matrix elements do not exhibit a strong hierarchy and there is a strong correlation between some of the mass matrix elements of up and down sector. A precision measurement of $\delta$ as well as small quark masses  would  have the potential to constrain the phase structure of the matrices  further.
 \end{abstract} 
 
\section{Introduction}
A convincing theory of fermion masses and mixings is one of the most salient omission in the current description of particle physics and at present it does not seem to be on the horizon. All of the charged fermion masses and mixing parameters in the quark sector are experimentally known to a very good precision \cite{pdg}. Even in the neutrino sector, remarkable experimental progress has resulted in advancement in our knowledge of the properties of neutrinos. Theoretical  attempts, on the other hand, are limited mostly to phenomenological approaches, wherein one goes for building phenomenological models having the potential for  predictions, in tune with the latest data, which later on become the guiding steps for more ambitious theories.

In the standard electroweak model, the masses of the charged fermions, weak mixing angles and the complex phase parameter causing CP violation enter as free parameters. Further
insights into the dynamics which determine the masses and mixing angles call for steps beyond the
electroweak standard model. In the absence of specific hints towards the underlying dynamics
one is invited to consider specific symmetry schemes which could reduce the number of free
parameters in the fermion sector and eventually provide hints towards further dynamical
details of the mass generation mechanism \cite{famsym}.

In this regard, one of the successful approach is the ``texture zero" approach, initiated implicitly by Weinberg \cite{weinberg} and explicitly by Fritzsch \cite{frz}, wherein certain entries of the mass matrix are exactly zero, such matrices are usually referred to as ``texture specific mass matrices". The language of texture zeros has proven to be very useful in establishing experimentally testable relations between the ratios of quark masses and the flavor mixing angles \cite{mmreview}.  Recently it has been shown \cite{shivali} that the present refined data of quark masses and mixing angles unambiguously rules out the original Fritzsch ansatz as well as all of its texture 6 zero variants. Further, also the present data rules out all possible texture 5 zero hermitian mass matrices \cite{5zero}. On the other hand, it has been observed over the years that  texture 4 zero mass matrices, similar to the original Fritzsch ansatz have always been compatible with the quark mixing data \cite{mmreview,4zero}. It has also been shown that out of all possible texture 4 zero mass matrices, compatible with weak basis transformations \cite{weakbasis}, there is a unique set of matrices along with its permutations  matrices which has the best possible compatibility with the quark mixing data. This is also borne out by a very recent extensive analysis of  of  texture 4 zero mass matrices \cite{aakriti}. This brings into focus  the issue whether it is possible to reduce the free parameters of these matrices so as to facilitate the task of model builders. The purpose of the present paper is to explore this possibility, particularly in relation to the phase parameters of these having implications for the CP violating parameters of the Cabibbo-Kobayashi-Maskawa (CKM) matrix.

The paper is organized as follows. In Section \ref{qmm}, while presenting the analysis of Fritzsch-Xing matrices with updated values of quark masses and the latest results of the CKM flavor mixing parameters, we also give the notations and basic expressions needed and general formalism adopted for the analysis. In Section \ref{mqmm}, we formulate texture 4 zero mass matrices with minimal parameters, thus enhancing the predictive power of these mass matrices. After confirming the validity of this set of mass matrices, we also discuss the hierarchy exhibited by mass matrix elements in our model. The details regarding the parameter space spanned by the free parameters have been presented and the correlations between these have been discussed. In the Section 
\ref{secpre}, we present the predictions of this set of mass matrices for the angles of the reference unitarity triangle (UT), Jarlskog's rephasing invariant parameter $J$ and CP violating phase $\delta$. In Section  \ref{sensitivity}, we study the relative dependence of CKM parameters on the quark masses. Finally, Section \ref{sum} summarizes our main conclusions.

\section{The Fritzsch-Xing Texture 4-zero Quark Mass matrices: An updated analysis} \label{qmm}
Before performing an updated and complete numerical analysis of the parameter space
of mass matrices with four texture zeros, let us briefly review the relations
between the parameters of mass matrices and the observable quantities.
For the three generations of quarks,  the quark mass terms are given as
\be 
{\overline u}_L M_u u_R  + {\overline d}_L M_d d_R\,,
 \label{mc} \ee
where $u_{L(R)}$ and  $d_{L(R)}$ are the left-handed (right-handed)
 quark fields and $M_u$ and  $M_d$ are the mass matrices for the  up sector 
$(u,c,t)$ and down sector $(d,s,b)$ respectively.
The mass matrices  can be  diagonalized  by 
the following bi-unitary transformations :
\be V_{u_L}^{\dagger}M_u V_{u_R} = M_u^{diag} \equiv {\rm Diag}\,
(m_u,m_c,m_t)\,;
\label{ou} \ee
\be V_{d_L}^{\dagger}M_d V_{d_R} = M_d^{diag} \equiv {\rm Diag}
\,(m_d,m_s,m_b)\,;
\label{od} \ee
 where $M_q^{diag}~(q=u,d)$ are real and diagonal, while $V_{q_L}$ and
  $V_{q_R}$ are unitary $3 \times 3$ matrices. The quantities $m_u, m_d$ etc. denote the
  eigenvalues of the mass matrices, i.e. the physical 
quark masses.
 The quark mixing matrix, connecting mass eigenstates to the weak eigenstates,
referred to as the Cabibbo-Kobayashi-Maskawa 
matrix \cite{ckm} is given as, 
\be  V_{\rm CKM} =  V_{u_L}^{\dagger} V_{d_L}\,=  \left( 
\ba {ccc}
V_{ud} & V_{us} & V_{ub} \\
V_{cd} & V_{cs} & V_{cb} \\
V_{td} & V_{ts} & V_{tb} \\ \ea 
\right)\,, \label{mix} \ee
 where $V_{ud}$ gives the amplitude of the process $u \rightarrow d +
W$ and similarly for the other elements. Thus, apart from the quark masses, the elements of $V_{\rm CKM}$ constitute the
low energy observable quantities through which one can get clues
about the structure of the mass matrices, which are completely arbitrary in Standard model.  The  $V_{\rm CKM}$, by definition, is a  unitary
matrix,  hence can be expressed in terms of
 three real angles and six phases. Out of
the six phases, five can be re-absorbed into the quark fields in
Eq.(\ref{mc}). Therefore, one is left with only
one non-trivial phase
which is responsible for CP violation in the SM.
There are several parametrizations of the  $V_{\rm CKM}$,
however the most commonly used
parametrization is the standard parametrization, also advocated  by PDG \cite{pdg},
\be V_{{\rm CKM}} =  
 \left( 
\arraycolsep=1.4pt
 \ba {ccc} c_{12} c_{13} & s_{12} c_{13} &
  s_{13}e^{-i\delta} \\
  -s_{12} c_{23} - c_{12} s_{23} s_{13}e^{i\delta} &
 c_{12} c_{23} - s_{12} s_{23}s_{13}e^{i\delta}
  & s_{23} c_{13} \\
  s_{12} s_{23} - c_{12} c_{23} s_{13}e^{i\delta} &
  - c_{12} s_{23} - s_{12}c_{23} s_{13}e^{i\delta} &
  c_{23} c_{13} \ea 
 \right),  \label{ckm1}  \ee 
  with $c_{ij}={\rm cos}\,\theta_{ij}$ and
   $s_{ij}={\rm sin}\,\theta_{ij}$.  In the PDG representation,  sin$\delta \neq 0$
implies the existence of  CP violation.

The Fritzsch-Xing quark mass matrices (texture 2-zero for $M_u$ and texture 2-zero for $M_d$) are presented below,
\begin{equation}M_u= \begin{pmatrix}
0&A_u&0\\A^*_u&D_u&B_u\\0&B_u^*&C_u
\end{pmatrix};~~
M_d= \begin{pmatrix}
0&A_d&0\\A_d^*&D_d&B_d\\0&B_d^*&C_d
\end{pmatrix},
\label{eq:ourMM}
\end{equation}
where $A_u=\lvert A_u\rvert e^{i \alpha_u}, \,A_d=\lvert A_d\rvert e^{i \alpha _d}, \,B_u=\lvert B_u\rvert e^{i \beta_u}, \,B_d=\lvert B_d\rvert e^{i \beta_d}$, such that the combined texture of quark sector is texture 4-zero.
  
\vskip 0.2cm
The diagonalizing transformations for these matrices can easily be obtained in terms of quark  masses and free parameters $D_u,\, D_d,\, \phi_1=\alpha_u-\alpha_d$ and $\phi_2=\beta_u-\beta_d.$ The details of the diagonalization equations for these mass matrices can be found in our earlier work \cite{diagon}. 
Thus, in total, there are 10 parameters in the above set of matrices to describe the six quarks masses and four mixing parameters.
 
Before discussing the details of our analysis, we present the latest quark masses and mixing data. Since the observed parameters for the CKM mixing matrix are fitted at the electroweak energy scale
 $\mu = m_Z$, we have taken the quark masses also at $m_Z$ energy scale \cite{Qmasses}, and are given in Table \ref{tabinp}.
 
\begin{table}  
\bc
\caption{Quark masses at $m_Z$ energy scale  \cite{Qmasses} and mixing parameters as given by  global fit results by PDG \cite{pdg}.}
\begin{tabular}{ll} \hline
 $m_u$ & $1.23\pm 0.21$ MeV \\
 $m_c $& $0.620 \pm 0.017 $ GeV \\
 $m_t$ &$168.26 \pm 0.75$ GeV  \\
$m_d $& $2.67 \pm 0.19$ MeV\\
$ m_s$ &$ 53.16 \pm 4.61$ MeV\\
$m_b$  & $2.839 \pm 0.026$ GeV\\
$\lvert V_{us}\rvert $ & $0.22650 \pm 0.00048$ \\
$\lvert V_{ub}\rvert $ & $0.00361^{+0.00011}_{-0.00009}$ \\
$\lvert V_{cb}\rvert $ & $0.04053^{+0.00083}_{-0.00061}$\\
$ \sin 2\beta $ & $0.699 \pm 0.017$ \\
 \hline
\end{tabular}\label{tabinp}
\ec
\end{table} 

For the mass matrices given in Eq. (\ref{eq:ourMM}), the CKM matrix can be obtained in terms of quark masses and free parameters $D_u, D_d, \phi_1$ and $\phi_2$. For the analysis, we use the quark masses given in Table \ref{tabinp} and scan their full ranges at 1$\sigma$, while the phases are scanned in the range [$-\pi,\pi$]. Thereafter, we calculate the CKM elements $\lvert V_{us}\rvert , \; \lvert V_{cb}\rvert , \lvert V_{ub}\rvert $ and angle $\beta$ of the UT. We keep the parameter sets, in which the value of each of these four observables is reproduced within the 1$\sigma$ interval of error-bars, as given by the global fit results by Particle Data Group (PDG) \cite{pdg} and presented  in Table \ref{tabinp}.
  
We choose these four mixing parameters, because their values have been determined to a very good degree of accuracy. The set of matrices given in Eq.(\ref{eq:ourMM}), successfully reproduces the three observed CKM elements and angle $\beta$ given in Table \ref{tabinp} in the parameter ranges given in Table \ref{tabpara}.

\begin{table}[h]
\begin{center}
\caption{Parameter ranges for the quark mass matrices given in Eq.(\ref{eq:ourMM}) that reproduce the mixing
parameters given in Table \ref{tabinp}.}
\begin{tabular}{cccc}
 \hline
$D_u\,(GeV)$~ & ~$D_d\,(GeV)$ &$\phi_1$& $\phi_2$ \\
$19.0-61.0$~ & ~$0.3-0.9$& $69.0\degree - 122.0\degree$ & $2.0\degree - 7.2\degree$ \\
\hline
\end{tabular}
\label{tabpara}
\end{center}
\end{table}

Angle $\beta$ refers to one of the inner angles of the CKM unitarity triangle described by the orthogonality relation $$ V_{ud}V_{ub}^{*} + V_{cd}V_{cb}^{*} + V_{td}V_{tb}^{*}
  =  0\,, $$ in the complex plane. The three
inner angles of this triangle are defined as,
 \beqn
  \alpha &\equiv &{\rm arg} \left[
-\frac{V_{td}V_{tb}^{*}}{V_{ud}V_{ub}^{*}} \right],\nonumber\\  
\beta &\equiv &{\rm arg} \left[
-\frac{V_{cd}V_{cb}^{*}}{V_{td}V_{tb}^{*}} \right],\nonumber\\  
\gamma &\equiv& {\rm arg} \left[
-\frac{V_{ud}V_{ub}^{*}}{V_{cd}V_{cb}^{*}} \right].   \label{angle} 
\eeqn  

  Another interesting measure of CP violation is the Jarlskog's rephasing invariant parameter $J$.
   The significance of ${J}$
     lies in the fact that all the CP violating effects in the SM
     are proportional to it as well as it is independent of the
     representation of CKM matrix. For hermitian quark mass matrices, $J$ can be obtained 
using the commutator of mass matrices   \cite{jarlskog}, for example,
\begin{equation} iC \equiv [M_u,M_d] \, , \end{equation}
 such that
 \beqn \textrm{Det} C &=&-2J(m_t-m_c)(m_c-m_u)(m_u-m_t) \nonum \\
 & & (m_b-m_s)(m_s-m_d)(m_d-m_b). \label{eq:detC} \eeqn
In parallel with the usage of commutators encountered in quantum mechanics, the commutator of hermitian mass matrices is a measure of the simultaneous diagonalization of the matrices, that is, its non zero value ensures mismatch in the diagonalization of the mass matrices, which in turn implies $CP$ violation.   Using Eq.(\ref{eq:detC}), one can easily express the rephasing invariant $J$, in terms of the mass matrices, $M_u$ and $M_d$.
  One can also express $J$ in terms of the $V_{\rm{CKM}}$ elements, as \be J \sum_{\gamma , l} \epsilon_{\alpha \beta \gamma}\epsilon_{j k l } =\textrm{Im}[V_{\alpha j} V_{\beta k} V_{\alpha k}^* V_{\beta j} ^*]. \ee
Using the standard parametrization, given in Eq.(\ref{ckm1}), $J$ can be expressed in terms of the CKM mixing angles, $\theta_{12}, \theta_{23} \text{ and } \theta_{13}$, and $CP$ violating phase, $\delta$, as
\begin{equation} J= s_{12}s_{23}s_{13}c_{12}c_{23}c_{13}^2 \sin\delta. \label{eq:deltaformula}\end{equation} 
Therefore,
\be 
\footnotesize{
\delta = \sin^{-1} \left( \frac{- \textrm{Det} C}{2(m_t-m_c)(m_c-m_u)(m_u-m_t)(m_b-m_s)(m_s-m_d)(m_d-m_b)  s_{12}s_{23}s_{13}c_{12}c_{23}c_{13}^2} \right).}
\label{eq:dfromJ}
\ee 

\noindent
Thus, the $CP$ violating phase $\delta$ can be obtained in a rephasing invariant manner using the commutator of mass matrices. 
In Table \ref{tabpr1}, we present the predictions of the mass matrices given in Eq. (\ref{eq:ourMM}) for the other two angles of the unitarity triangle i.e. $\alpha$ and $\gamma$ and for  $J$ and $\delta$.  From the Table, we observe that the ranges of $ \alpha,\,\gamma,\,J$ and $\delta$ 
are quite wide, however these include their ranges obtained from the global analyses  and the experimental limits \cite{pdg}.
 
\begin{table}[h]
\bc
\caption{Predictions of the quark mass matrices given in Eq.(\ref{eq:ourMM}).
}
    \begin{tabular}{ccc}
\hline
$ \alpha$ & $66.2\degree - 113.7\degree$  \\
$\gamma$ & $44.7\degree -  92.3\degree$\\
$J$ & $(2.30 - 3.39)\times 10^{-5}$\\
$\delta$ &  $45.0\degree - 90.0\degree$\\ \hline
\end{tabular}\label{tabpr1}
 \ec
\end{table}

\section{Texture 4-zero Quark Mass matrices with minimal parameters} \label{mqmm}
 A figure of merit for any mass matrix model is its predictability power, which is maximal when a minimal number of free parameters are introduced. Using simplicity and the requirement of maximal predictability as guiding principles, we are led to consider if we can reduce the number of parameters further in the set of mass matrices given in Eq.(\ref{eq:ourMM}). To this end, we have tried various combinations, for example, we considered some of the elements as zeros, e.g. one can put (2,2) or (2,3) or (3,3) element as zero, however this does not reproduce CKM matrix within its limits. Getting rid of one of the phases $\phi_1$ or $\phi_2$, also does not lead to reproduction of compatible CKM matrix elements. 

\begin{figure}[h]
 \centering
 \includegraphics[width=.8\textwidth] {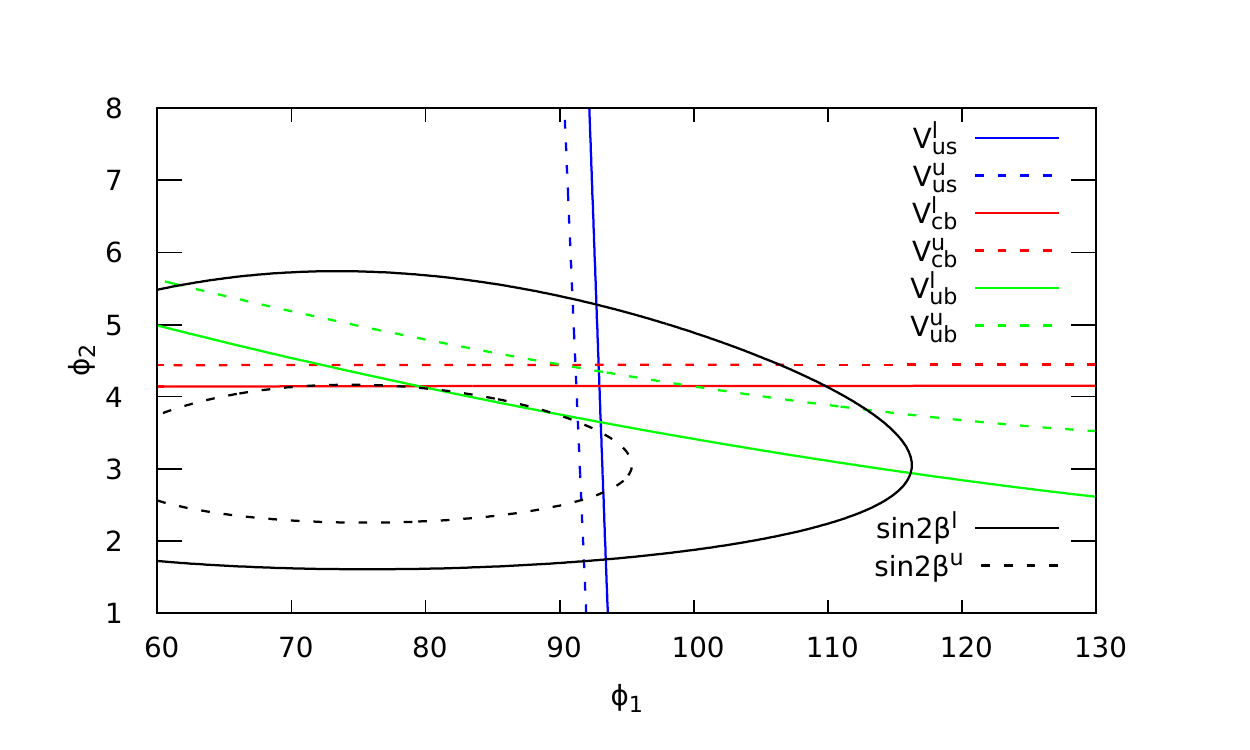} 
 \caption{Contours of $\lvert V_{us}\rvert , \lvert V_{cb}\rvert , \lvert V_{ub}\rvert $ and $\sin2\beta$ in $\phi_1-\phi_2$ plane, for the mean values of quark masses and central values of allowed ranges of $D_u$ and $D_d$. The solid lines are for the lower limit of the ranges mentioned in the Table \ref{tabinp}, while the dashed lines are for the upper limit.}
 \label{fig:contours}
 \end{figure}

 In Fig. \ref{fig:contours}, we have plotted contours of $\lvert V_{us}\rvert , \lvert V_{cb}\rvert , \lvert V_{ub}\rvert $ and $\sin2\beta$ in $\phi_1-\phi_2$ plane for the mean values of quark masses and central values of allowed ranges of $D_u$ and $D_d$ given in Table \ref{tabpara}. The solid lines are for the lower limit of the ranges mentioned in the Table \ref{tabinp}, while the dashed lines are for the upper limit. The figure reveals several interesting points. For example, it   suggests that $\lvert V_{us}\rvert $ is not sensitive to variation in $\phi_2$, while it is very sensitive to variation in $\phi_1$. On the other hand, $\lvert V_{cb}\rvert $ is not sensitive to variation in $\phi_1$, while it is very sensitive to variation in $\phi_2$. $\lvert V_{ub}\rvert $ and $\sin 2\beta$ are sensitive to both $\phi_1$ and $\phi_2$. However, the most important observation is that angle $\phi_1$ is close to $90\degree$, while angle $\phi_2$ is very close to $0\degree$, where these strong constraints are majorly put by $\lvert V_{us}\rvert $ and $\lvert V_{cb}\rvert $, respectively. We also find that a non zero value of $\phi_2$ is very crucial to reproduce angle $\beta$. A quick look at Fig. \ref{fig:contours} motivates us to consider the possibility $\phi_1 = \phi_2 + 90\degree$, which we have investigated in detail below. As discussed earlier, our endeavor is to reduce the number of free parameters in mass matrices and hence we now present texture 2-zero parallel structure for the up and down quark mass matrices, with one less parameter than the matrices given in Eq. (\ref{eq:ourMM}), for example 
\beqn
& &M_u= \begin{pmatrix}
0&\lvert A_u\rvert e^{i (\beta_u +\pi/4)}&0\\\lvert A_u\rvert e^{-i (\beta_u +\pi/4)}&D_u&\lvert B_u\rvert e^{i \beta_u}\\0&\lvert B_u\rvert e^{-i \beta_u}&C_u 
\end{pmatrix};~~ \nonumber \\
& & M_d= \begin{pmatrix}
0&\lvert A_d\rvert e^{i (\beta_d -\pi/4)}&0\\\lvert A_d\rvert e^{-i (\beta _d -\pi/4)}&D_d&\lvert B_d\rvert e^{i \beta_d}\\0&\lvert B_d\rvert e^{-i \beta_d}&C_d
\end{pmatrix}.
\label{eq:newMM}
\eeqn
The mixing matrix corresponding to above mass matrices can be obtained in terms of the quark masses and free parameters 
  $D_u,\, D_d$ and $\phi_2=\beta_u-\beta_d.$  
Thus, in our proposed model we consider only one phase, i.e. $\phi_2$ to reproduce the CKM elements and angle $\beta$ of the UT.
As before, we scan the parameter space of $D_u, D_d$ and $\phi_2$ and keep the parameter ranges, in which the value of each observable is reproduced within the $1\sigma$ interval. We find that this simple \textit{ansatz} is able to reproduce the CKM elements, angles of UT, $J$ and $\delta$ well within their ranges given in PDG, in the parameter ranges given in Table \ref{tab:res3}. \par
\begin{figure*} 
 \centering
 \includegraphics[width=.75\textwidth] {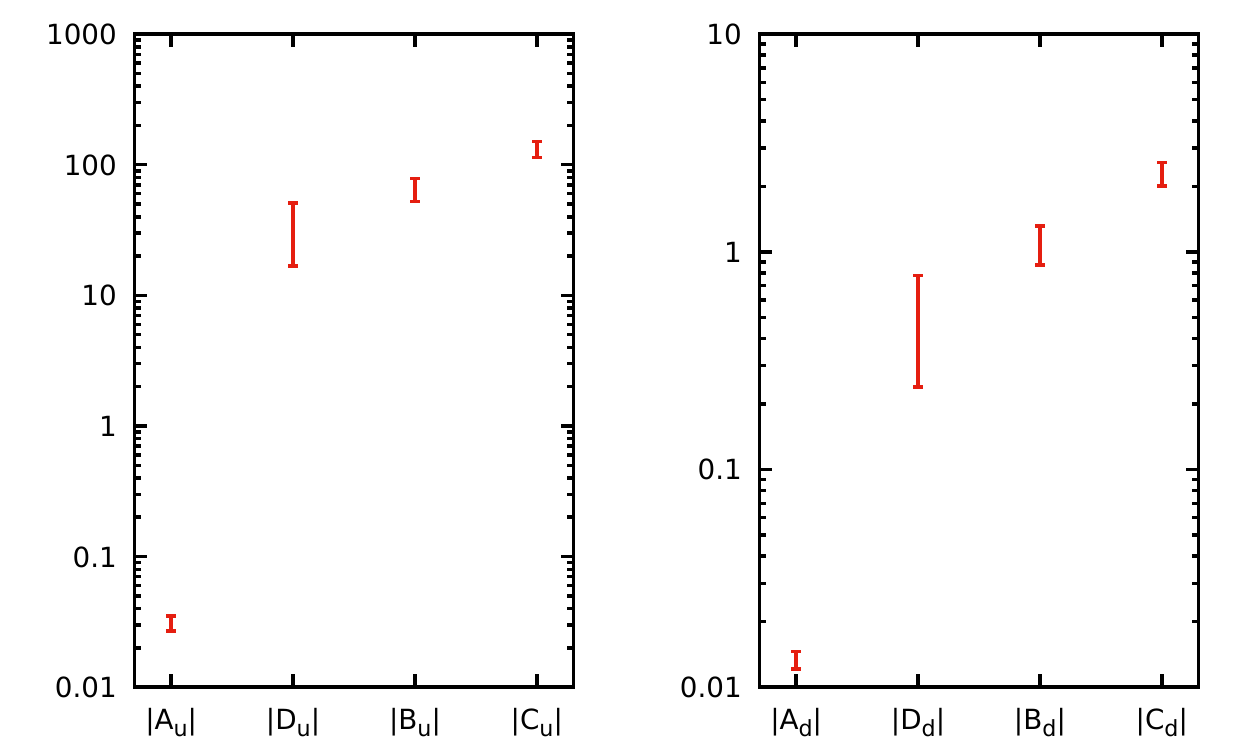} 
 \caption{The ranges of mass matrix elements for both $M_u$ and $M_d$, depicting the hierarchy observed.}
 \label{fighier}
 \end{figure*}

\begin{table}[h]
\centering
\caption{Parameter ranges for the mass matrices given in Eq. (\ref{eq:newMM}) that reproduce the mixing
parameters given in Table \ref{tabinp}.}
\begin{tabular}{ccc}
\hline
$D_u\,(GeV)$~~ & ~~$D_d\,(GeV)$ & $\phi_2 $\\
$17.0-51.5$ & $0.25-0.75$ & $2.5\degree - 7.7\degree$\\
\hline
\end{tabular}
\label{tab:res3}
\end{table}
Once the allowed ranges of these three parameters are obtained, we can evaluate various quantities of the CKM paradigm. We begin with presenting the allowed ranges of real parts of the mass matrices $M_u \text{ and } M_d$, 
\begin{eqnarray}
M_u^r&=&\begin{pmatrix}
0&0.027 - 0.035&0\\
0.027 - 0.035 & 17-51.5 & 51.47 - 78.26\\
0 & 51.47 - 78.26 & 115.88 -150.91
\end{pmatrix};~~ \nonumber \\
M_d^r&=&\begin{pmatrix}
0 & 0.012 - 0.015 & 0\\
0.012 - 0.015 & 0.25 - 0.75 & 0.88 - 1.30 \\
0 & 0.88- 1.30 & 2.02 - 2.56
\end{pmatrix}.
\label{eq:MMnum}
\end{eqnarray}
In the CKM paradigm, we notice several points of importance. First, the quark masses are strongly hierarchical, $m_{u,d }\ll m_{c,s} \ll m_{t,b}$, with the hierarchy of `up' sector being much stronger. Second, the CKM matrix elements also exhibit a strong hierarchy, i.e. $\lvert V_{ub}\rvert  < \lvert V_{td}\rvert \ll \lvert V_{ts}\rvert  < \lvert V_{cb}\rvert \ll \lvert V_{cd}\rvert  <\lvert V_{us}\rvert  < \lvert V_{cs}\rvert  < \lvert V_{ud}\rvert  < \lvert V_{tb}\rvert $, indicating the suppression of off-diagonal elements in comparison to diagonal ones. Therefore, we expect our analysis to manifest these strong hierarchies as well, i.e., the mass matrix elements are expected to follow the hierarchy $(3, 3) \gg (2, 3) \gg (2, 2) \gg (1, 2)$. However, our results in Eq.(\ref{eq:MMnum}) show that the hierarchy followed in both up and down sector is $(3, 3) \gtrsim (2, 3)\gtrsim (2, 2) \gg (1, 2)$, i.e. except for (1,2) element, the hierarchy observed between the mass matrix elements is weak. 
Fig. \ref{fighier} clearly depicts the hierarchy followed by mass matrix elements, wherein 
we show the obtained ranges of the mass matrix elements.
Various authors have also considered ‘weakly’ hierarchical mass matrices to reproduce ‘strongly’ hierarchical mixing angles \cite{our4zero}.  
 For instance, in Reference \cite{Xing2004}, Xing \textit{et al.} find $\lvert B_i\rvert /C_i \sim D_i/\lvert B_i\rvert  \sim 0.24$ for the case of both up and down  mass matrices. However, in Reference \cite{1501.06346}, Xing and Zhou find a new part of the parameter space, corresponding to $C_i \sim \lvert B_i\rvert  \sim D_i$. Both of these possibilities are covered by the set of mass matrices formulated by us in Eq. (\ref{eq:newMM}). To understand the reason to realize strong hierarchy in the elements of CKM matrix obtained from the weakly hierarchical mass matrix elements, we present the approximate expressions for the  off diagonal  CKM element  $V_{cb}$ below. We have used the approximation $m_{u,d }\ll m_{c,s} \ll m_{t,b}$ to arrive at the following:
\be 
  V_{cb}  \approx  -\sqrt{\frac{(m_t-D_u)(D_d+m_s)}{m_t m_b}}+e^{i \phi_2}\sqrt{
\frac{(D_u+m_c)(m_b-D_d)}{m_t m_b}}\,.\ee 
Numerical analysis shows that (i) $\phi_2$ being very small and the assumption $\phi_1 = \phi_2 + 90\degree$ as well as (ii) the lower right corner of the $M_u$ and $M_d$ being weakly hierarchical (as presented in Eq. (\ref{eq:MMnum})) ensure that $\lvert V_{cb}\rvert$ is sufficiently small to be in its experimental range. The free parameters   $D_u$ and $D_d$ of our mass model, being in the range  given in Eq. (\ref{eq:MMnum}) have a profound role in reproducing $\lvert V_{cb}\rvert$ and other off diagonal elements of CKM matrix. If the mass matrices given in Eq. (\ref{eq:newMM}) are taken to be  strongly  hierarchical, the off diagonal elements of CKM matrix turn out to be larger than  their experimental ranges. This issue has also been extensively discussed in a recent paper by Fritzsch et al. \cite{frzrev}, wherein they have studied the correlations between quark mass and flavor mixing hierarchies for Texture 4-zero Fritzsch-Xing quark mass matrices.

\begin{figure*} 
 \centering
   \includegraphics [width=.8\textwidth]{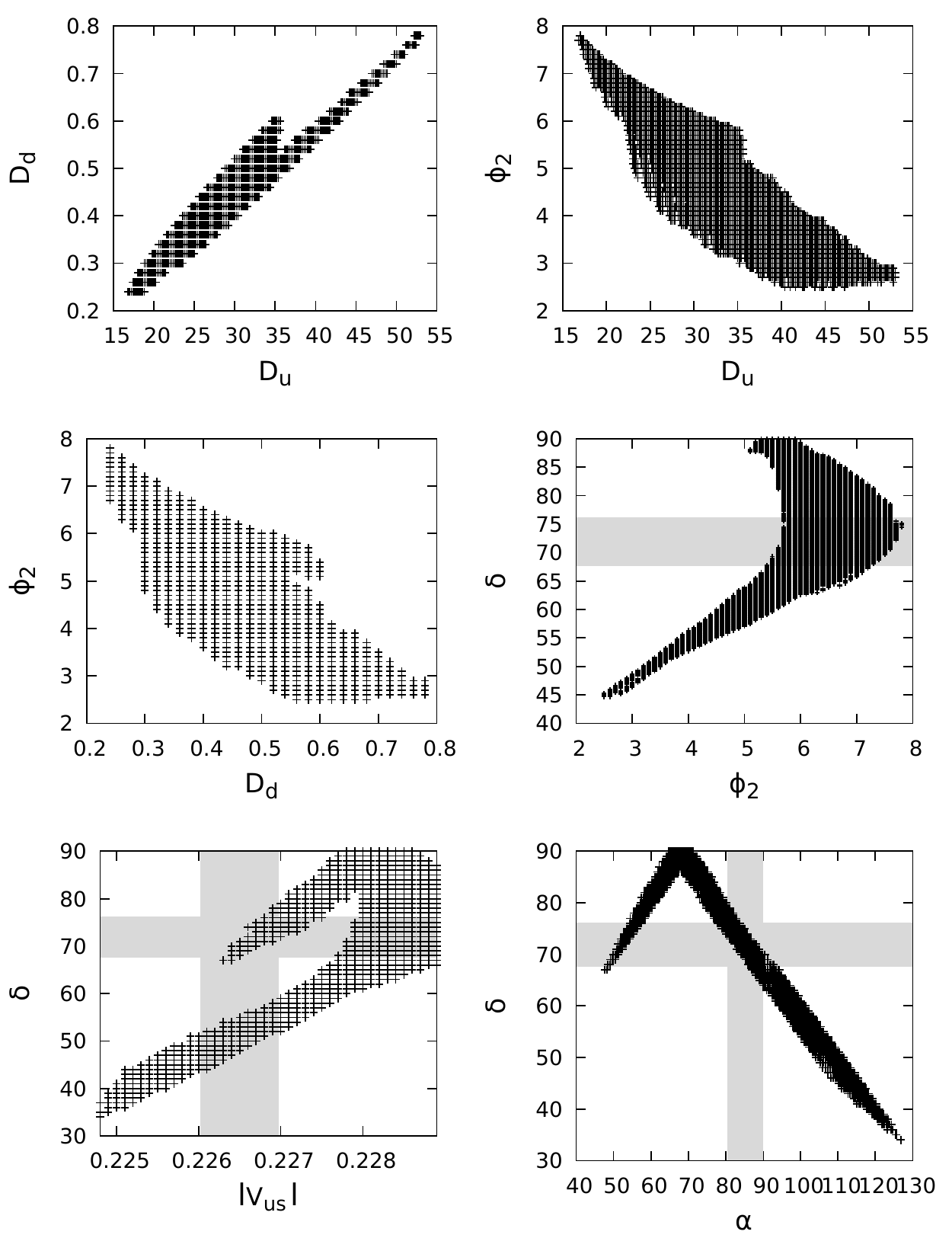}  
 \caption{The top and middle panels show the parameter space spanned by the free parameters of our mass matrices model given in Eq. (\ref{eq:newMM}), while the lowest panel displays the
correlations between   $\lvert V_{us} \rvert$ and $\delta$ and between $\alpha$ and $\delta$.}
 \label{figcorr}
 \end{figure*}

Apart from the hierarchy between the elements of $M_u$ or $M_d$, it is also desirable to discuss the correlations amongst the free parameters of $M_u$ and $M_d$. In Fig. \ref{figcorr}, we present the complete parameter space spanned by the free parameters, $D_u, D_d \text{ and } \phi_2$ in our model. We find that there is a strong correlation between mass matrix element (2,2) of up and down sector, i.e., $D_u$ and $D_d$. Further, the correlation between $D_u$ and $\phi_2$ and between $D_d$ and $\phi_2$ is very similar, hinting towards an underlying symmetry in up and down sectors. 
In the middle right panel, we have plotted the parameter space available to $\delta$ with respect to $\phi_2$. We see a very strong correlation between $\delta$ and $\phi_2$. The shaded region corresponds to the limits on $\delta$ by the present global range, i.e. $\delta = (72.1^{+4.1}_{-4.5})\degree$ \cite{pdg}. We find that  the global range of $\delta$  is compatible with $\phi_2$ to be around $6\degree$. 
Furthermore, the scatter plots in the lowest panel of Fig.  \ref{figcorr}, show very interesting correlations between some of the flavor mixing observables.
We find that  $\delta$ is correlated to $\lvert V_{us} \rvert$, clearly a result of lesser number of free parameters in our model due to the added assumption of 
$\phi_1 = \phi_2 + 90\degree$. These correlations might be useful to   explore the underlying correlations between the quark mass spectrum and the flavor mixing pattern. Also, there is an  obvious linear correlation between  $\alpha$ and $\delta$ owing to the the unitarity triangle relation $\alpha+\beta+\gamma=\pi$. 
The grey shaded regions in these plots indicate the 1$\sigma$ experimental ranges \cite{pdg} for these observables.

\begin{figure*}  \centering
 \includegraphics [width=1.0\textwidth]  {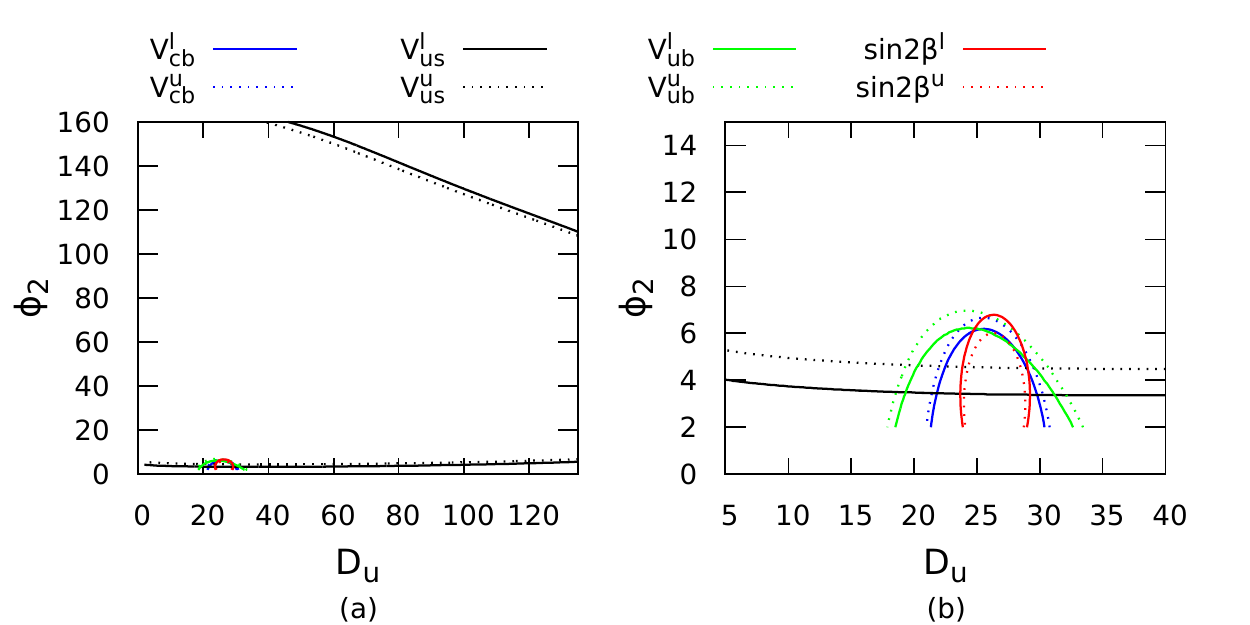}  
 \caption{(a) Contours of $\lvert V_{us}\rvert , \lvert V_{cb}\rvert , \lvert V_{ub}\rvert $ and $\sin2\beta$ in $D_u$ and $\phi_2$ plane, for mean values of quark masses and $D_d = 0.4$, for the matrices given in Eq.(\ref{eq:newMM}). The solid lines are for the lower limit of the range mentioned in the Table \ref{tabinp}, while the dashed lines are for the upper limit. (b) The zoomed in region of (a) in the allowed range.}
 \label{figcontnew}
 \end{figure*}

 In Fig. \ref{figcontnew}(a), we have plotted the contours of $\lvert V_{us}\rvert , \lvert V_{cb}\rvert , \lvert V_{ub}\rvert $ and $\sin2\beta$ in $D_u$ - $\phi_2$ plane, for mean values of quark masses and $D_d =0.4$. The Fig. \ref{figcontnew}(b) depicts the zoomed in region of the same figure for better clarity. We infer that the precision determination of $\sin2\beta$ (red lines) and $\lvert V_{cb}\rvert $ (blue lines) will significantly restrict the allowed range of $D_u$. However, $\lvert V_{us}\rvert $ (black lines) does not put strong constraints on $\phi_2$ or $D_u$. We also find that CKM element $\lvert V_{ub}\rvert $ (green lines) loosely restricts $D_u$. The phase $\phi_2$ is restricted by  $\lvert V_{cb}\rvert ,\,\lvert V_{ub}\rvert $ and  $\sin2\beta$. These conclusions remain valid, even when the value of $D_d$ is varied in its allowed range. Thus, a more precise determination of $\sin 2\beta$ and $V_{CKM}$ elements will result in an even more restricted range of $\phi_2$, which would further yield a more precise evaluation of $CP$ violating phase $\delta$ as $\phi_2$ has interesting implications for the $CP$ violating phase $\delta$, as emphasized above.

\subsection{Predictions for CKM matrix, UT angles, $J$ and $\delta$}\label{secpre}
The mass matrix we formulated in Eq.(\ref{eq:newMM}), allows us to calculate the ranges of elements of the CKM mixing matrix using the methodology discussed earlier. the CKM matrix obtained is given as
\begin{eqnarray}
V_{CKM}=\begin{pmatrix}
0.9739 - 0.9741 & 0.2260 - 0.2270 & 0.00352 - 0.00372 \\
0.2258 - 0.2269 & 0.9731- 0.9733 & 0.0399- 0.0414\\
0.0070 - 0.0099 & 0.03889 - 0.04081 & 0.9991- 0.9992
\end{pmatrix}. 
\end{eqnarray}
The above ranges are well consistent with the current global fit by PDG \cite{pdg}. The other two interior angles, $\alpha$ and $\gamma$, of the UT  can be evaluated from Eq.(\ref{angle}) to be
\be
\alpha =(66.4 - 113.6)\degree , \;
\gamma = (44.9 - 92.1)\degree , 
\ee
while the experimental value of $\beta$ was used as a constraint to check the viability of mass matrices. In order to evaluate Jarlskog's invariant rephasing parameter $J$ and $CP$ violating phase $\delta$, we use Eqs. (\ref{eq:detC}) and (\ref{eq:dfromJ}) to obtain
\be
J = (2.30 - 3.39) \times 10^{-5}, \quad \delta = (44.9 - 92.1)\degree. 
\ee

The above ranges are quite wide and include the ranges given by the global fit analysis and experiments \cite{pdg}. Similar results for $CP$ violating phase and other CKM parameters have been obtained in the literature. For example, in Reference \cite{2002.09152}, authors obtain $\delta = (87.9) \degree$ as best fit value for the case of a 4-zero textures having the interplay of $\mu - \tau$ reflection symmetry. Similarly in Reference \cite{1912.07573}, authors present five different  viable models of texture zeros originating from modular symmetry and obtain the best fit value of $\delta= (69.2 \pm 3.1)\degree$. 
Thus, our model, despite having only one phase parameter, is not only consistent with other such analyses, but also gives an excellent fit to most of the CKM parameters, which is phenomenologically very encouraging.  
\begin{figure*}[t!] 
 \includegraphics[width=1.\textwidth]{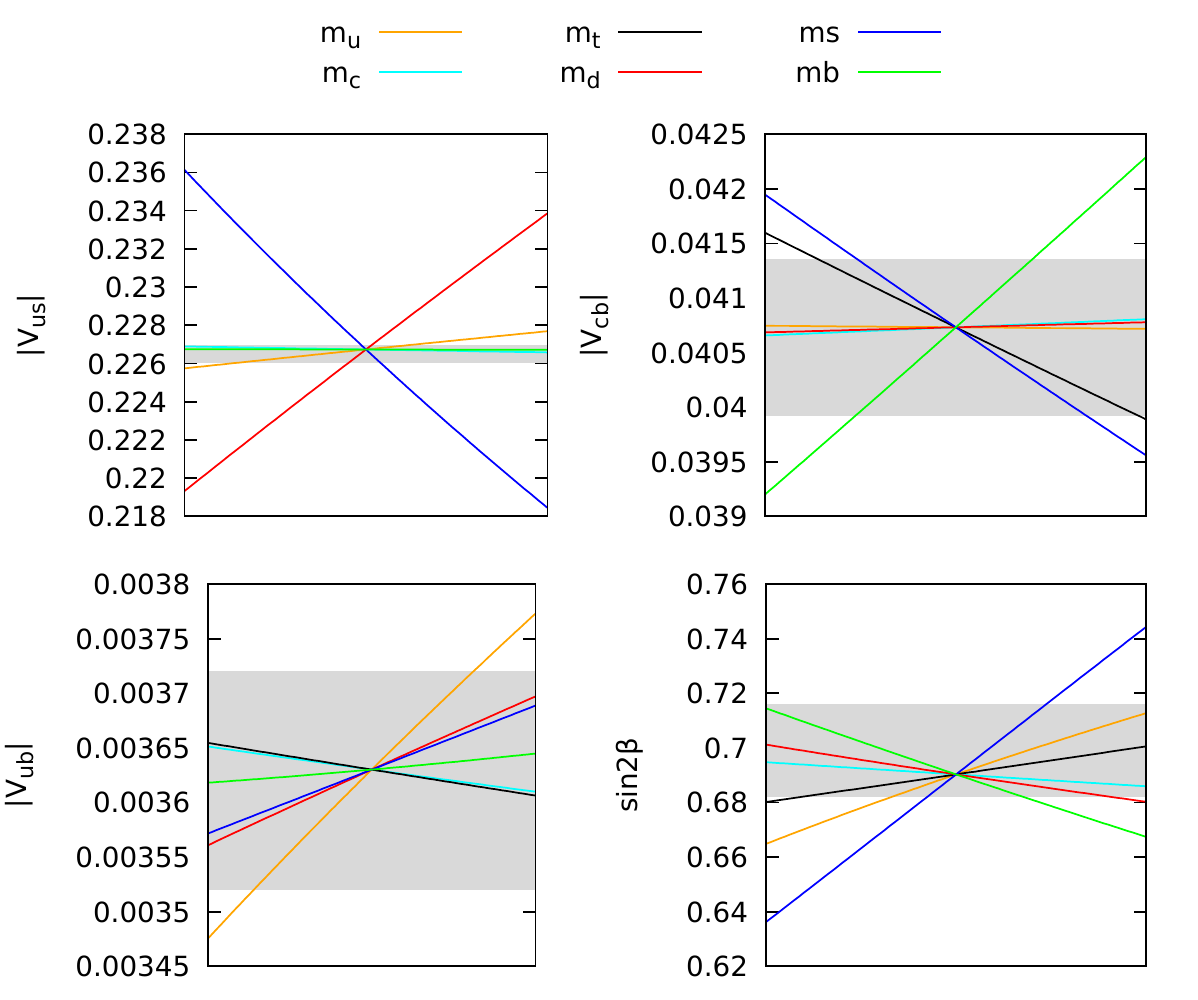}
 \caption{Dependence of $\lvert V_{us}\rvert $, $\lvert V_{ub}\rvert $, $\lvert V_{cb}\rvert $ and $\beta$ on quark masses $m_u$(orange), $m_c$(cyan), $m_t$(black), $m_d$(red), $m_s$(blue), $m_b$(green).The x-axis of each plot corresponds to the 1$\sigma$ variation in the quark masses, while the shaded region   corresponds to the experimental value of that particular mixing parameter at 1$\sigma$, as given in the Table \ref{tabinp}. }
 \label{figsens}
 \end{figure*}
\subsection{Sensitivity of CKM elements on quark masses}\label{sensitivity}
To understand the role of precision determination of the quark masses for the mass matrix model formulated by us, given in Eq. (\ref{eq:newMM}), we have shown the variation of $\lvert V_{us}\rvert $, $\lvert V_{ub}\rvert $, $\lvert V_{cb}\rvert $ and $\sin 2\beta$ with respect to the quark masses in Fig. \ref{figsens}.

The x-axis of each plot corresponds to the variation in the quark masses  $m_u,\, m_c,\, m_t,\, m_d,\, m_s$ and $m_b$, shown respectively by orange, cyan, black, red, blue and green lines. Each quark mass is varied from its minimum to maximum value in the $1\sigma$ range given in Table \ref{tabinp}. The shaded region in each plot corresponds to the experimental value of that particular mixing parameter at 1$\sigma$, as given in the Table \ref{tabinp}. A quick look at the Fig. \ref{figsens} indicates that $\lvert V_{us}\rvert $ is most sensitive to the variation in the masses $m_d \text{ and } m_s$, while $\lvert V_{ub}\rvert $ shows maximum sensitivity to $m_u$ along with $m_d$ and $m_s$. Similarly, $\lvert V_{cb}\rvert $ and $\beta$ are most sensitive to variation in $m_s$ and $m_b$. Since the measurement of light quark masses are not as precise as measurement of heavy quarks, therefore, further refinement in the measurement of these quark masses and CKM parameters will help in eliminating many of the quark mass models. 

\section{Summary and conclusion}\label{sum}
Motivated by the extensively used texture 4 zero Fritzsch-Xing  quark mass matrices, which are very successful in describing the CKM phenomenology, we formulate these with minimum possible free parameters. In the proposed structure, these quark mass matrices contain only one free phase compared with the usually considered  two phases. We study the implications of this particular set of texture 4 zero mass matrices  for the CKM phenomenology, particularly for the CP violation parameters,  $J$       and corresponding phase $\delta$. We find that these matrices  very well reproduce not only the CKM matrix but also the  angles of the UT,  $J$ and  $\delta$   well within their experimental and global fit ranges. We observe that except for the element (1,2), the matrix elements in both up and down sector follow a weak hierarchy, for example,  $(3, 3) \gtrsim (2, 3)\gtrsim (2, 2) \gg (1, 2)$. We find that there is a strong correlation between mass matrix element (2,2) of up and down sector, i.e.,  $D_u$ and $D_d$. Further, the correlation between $D_u$ and $\phi_2$ and between $D_d$ and $\phi_2$ is very similar, hinting towards an underlying symmetry in up and down sectors. We also infer that $\phi_2$ is strongly restricted by $\lvert V_{us}\rvert $ while $D_u$ and $D_d$ are primarily restricted by the limits $\sin 2\beta$ and $\lvert V_{cb}\rvert $. Due to a very strong correlation between the $CP$ violating phase $\delta$ and $\phi_2$, a precise measurement of $\delta$ will constrain the phase structure of the model further. The investigation of the sensitivity of various CKM elements on quark masses shows that a precision measurement of quark masses has the potential to constrain the model to a good extent.  

In conclusion, it can be said that even in the absence of  theoretical justification,  mass matrices  with fewer parameters are  not only simple, but also more predictive and serve as a useful guide for model building and calls for a theoretical investigations  to understand the underlying symmetries. Further, it remains to be seen whether this texture can also be applied to the lepton sector. 

\section*{Acknowledgements}
NA and MG would like to thank the Chairman, Department of Physics, Panjab University, Chandigarh  for providing facilities to work.
MK and MR would like to thank the Director, UIET,  Panjab University, Chandigarh  for providing facilities to work.  MK also acknowledges the financial support provided by Department of Science and Technology (DST), Govt. of India (Grant No. DST/INSPIRE Fellowship/2017/IF170125).

\end{document}